\documentclass[12pt,preprint]{aastex}

\begin{document}

\title{The beginning of the Universe}

\author{Stanis\l aw Bajtlik \altaffilmark{1}  }

 \affil{$^{1}$ Nicolaus Copernicus Astronomical Center,
       Bartycka 18, 00-716 Warszawa, Poland\\
       {\tt bajtlik@camk.edu.pl}}

\begin{abstract}
In this brief essay we consider the concept of 
beginning in cosmology. It can be applied to the Universe as a whole, 
as well as to various epochs in the cosmic history and major structures 
as we can see them today.
\end{abstract}

~\\

One of the most astonishing discoveries of the 20th Century was Hubble's 
discovery of the expansion of the Universe. Since time immemorial, people had 
shared a view of space and time as unchanging and independent of matter. Such a 
fixed space with a river of time, both unaffected by physical bodies, had been 
the stage for Newtonian mechanics. However, Hubble's discovery in the 
nineteen-twenties had come at the right time. In 1916, Einstein had published 
his General Theory of Relativity, and few years later Friedmann had found 
solutions of Einstein's equations corresponding to an expanding Universe. Cosmic 
space and time became related to matter; the geometrical properties of space and 
its temporal evolution became inseparable from the material content of the 
Universe and its large scale structure.

Cosmic expansion is of tremendous importance for studies of the BEGINNING of the 
Universe and the formation of the bodies and structures we observe. Since cosmic 
space expands, it follows that everything was closer together many epochs ago. 
As all cosmic objects have some fixed physical size, it is clear that there must 
have been some epoch in cosmic history, eons ago, when it was not possible for 
galaxies to exist as independent, individual structures. At earlier epochs, not 
even stars could exist; earlier still, even individual atoms would not fit into 
the much smaller spatial volumes; before that, even atomic nuclei would have 
been too large, and so on.

The concept of the BEGINNING of the Universe as a specific event in the past is 
deeply rooted in many religions and was discussed by the ancient philosophers. 
However, Hubble's discovery suggested something profoundly different. In 
mythology, the BEGINNING of the Universe was a one-time event, resulting in a 
world "ready to use", similar to what we see today. The expansion of space, 
however, proves that the Universe must have evolved, since it was impossible for 
organised structures (such as atoms, molecules, planets, stars and galaxies) to 
exist from the very BEGINNING.

Another profound consequence of the Universe having a BEGINNING is that, due to 
its finite age and the finite maximum speed at which energy, information or 
interaction can be transferred, (i.e., the speed of light), the observable, 
causally-connected part of the Universe must be finite, regardless of the 
overall structure or geometry of space. This finite volume is surrounded by the 
cosmic horizon. Nothing from behind the horizon can have influenced anything 
which we can see or experience.

Reversing (as a thought experiment) cosmic expansion and extrapolating backward 
in time, we can estimate the age of the Cosmos from its current expansion rate. 
The age of the Universe is the time which has passed since space and matter 
started to expand as a result of the Big Bang the BEGINNING. It is not an easy 
task, as the expansion is constantly being slowed by the combined self-gravity 
of all matter in the Universe. To know the rate of this deceleration we need to 
know the amount and form of matter in the Cosmos. The best present-day estimate 
for the age of the Universe is $13.7\times10^9$years. How far back in time we 
extrapolate is really a matter of taste. As we go closer and closer to the Big 
Bang, the physical conditions in the Universe were more and more extreme. It is 
clear that, at some point, the physical theories we know must break down. 
However, we can safely discuss processes in the primeval cosmic plasma a few 
seconds after the Big Bang. The conditions there were not that different from 
what we can study in nuclear reactors today. Even a mere fraction of a second 
after the BEGINNING, the physical conditions in the Universe (i.e. the density 
and temperature) are still familiar, as they are similar to what we can produce 
in particle accelerators.

There is, however, a definite limit to the extrapolation of known physics back 
to the BEGINNING of the cosmic history. Approximately 10-43 seconds after the 
Big Bang, at the so-called Planck time, the mass density of the Universe is 
thought to approach roughly $10^{93}$\, g\,cm$^{-3}$. Contained within the causally connected 
part of the Universe is then a Planck mass, roughly $10^{-5}$\,g. An object of such 
mass would be a quantum black hole, with an event horizon close to both its own 
Compton length (the distance over which a particle is quantum mechanically 
"fuzzy") and the size of the cosmic horizon at the Planck time. Under such 
extreme conditions, space-time cannot be treated as a classical continuum and 
must be given a quantum interpretation.

We still do not understand the properties of the Universe at the very BEGINNING. 
No existing theory can describe the initial singularity and no theory can 
explain why the Big Bang had to happen. Many researchers put much hope in the 
formulation of a so called  final theory  or  theory of everything . Some of 
them believe that string theory may be the best starting point for the 
formulation of such a theory.

The BEGINNING of the Universe is only the ultimate starting point in the history 
of cosmic "beginnings". A series of early phase transitions created the initial 
density fluctuations, which later served as the seeds of cosmic 
structure-superclusters, clusters and galaxies. It was the BEGINNING of 
structure formation. In inflationary models, the universe starts with very close 
to zero nett baryon number; at the BEGINNING, there was as much matter as 
antimatter. An excess of matter can only be generated from this initial 
situation after inflation under the following conditions: the baryon number is 
violated; the C (charge) and CP (charge-parity) symmetries are violated; the 
universe was not in thermal equilibrium for a period during bariogenesis. As the 
CP symmetry violation is known from laboratory physics, we believe that we 
understand why the Universe is made of matter rather then a mixture of matter 
and antimatter. All this happened in a tiny fraction of a second after the Big 
Bang.

We are made of chemical elements. Our bodies and everything around us is made of 
baryonic "stuff", organized in molecules which are themselves composed of atoms. 
The atomic form of matter was created in several processes during cosmic 
evolution. The lightest isotopes, those of hydrogen, helium and lithium, were 
created during the first three minutes after the BEGINNING. Later it was too 
cold and the density was too low for hierarchical build up of more complex 
atomic nuclei. The process of primordial nucleosynthesis was interrupted and 
resumed only after the first stars had formed. To explain the observed 
abundances of the light elements in the Universe, it was postulated that, at the 
BEGINNING, the Universe must have been hot. Matter and radiation were in thermal 
equilibrium. The existence of the remnant radiation, which today has cooled to 
just $2.7$\,Kelvin, was predicted. The discovery of this relict radiation in the 
microwave domain in the nineteen-sixties was a major turning point in the 
history of cosmology. The expansion of the Universe, the abundances of the light 
elements, and the existence and properties of the microwave relict radiation are 
the fundamental observational facts supporting the Big Bang scenario.

The BEGINNING of the formation of heavier chemical elements started in stars 
just few hundred million years after the Big Bang. This process has been 
continuing throughout the whole of cosmic history, altering the chemical content 
of galaxies. Because looking deep into space we are looking into the remote past 
(due to the finite speed of light), we can actually observe the process of 
structure formation and the chemical evolution of galaxies.

Galaxies and clusters of galaxies formed from the initial density fluctuations 
as a result of gravitational instability. They came into existence in the 
Universe soon after the Big Bang, less than $10^9$ years after the BEGINNING. Most 
stars were formed when the Universe was five times younger then today.

In the last ten years we have discovered planets beyond our Solar System. Today 
we know of more than one hundred extrasolar planets. We observe protoplanetary 
disks and we are becoming increasingly convinced that planet formation is a 
natural part of the process of stellar formation, and that planets should not be 
uncommon in the Universe. The estimated age of the Earth is only slightly less 
than the age of the Sun. Our planet, as well as the Sun and other planets in the 
Solar System, was formed from the same protostellar nebula, some $4.5\times10^9$\, years 
ago. Thus the BEGINNING of our planetary system and our own planet took place 
when the Universe was approximately three times younger than it is today.

Life started on the Earth soon after the planet cooled enough for complex 
molecules to form. The BEGINNING of life is estimated to have occurred about 
$3.5\times10^9$ years ago.

Scientists from the Copernicus Astronomical Center in Warsaw have taken an 
active role in this process of revealing cosmic evolution. The process of 
stellar evolution was studied (among others) by Professors Bohdan Paczy\'nski (now 
at Princeton University) and J\'ozef Smak. Recently, Professor Janusz Kaluzny 
measured the age of the one of oldest known stars in the Universe. Professor 
Roman Juszkiewicz has contributed to the understanding of the fluctuations in 
the relict microwave background, galaxy formation and created original methods 
of measuring the mean density of matter in the Universe. The author of this 
article has studied clouds of intergalactic matter and has estimated the 
intensity of the ionising ultraviolet background radiation, which played an 
important role in the structure formation epoch. Krzysztof G\'orski (now at JPL, 
Pasadena) and Rados{\l}aw Stompor (now at UC, Berkeley) have made important 
contributions into the investigation of the relict microwave radiation. Ewa 
{\L}okas and Micha{\l} Chodorowski contributed to the theory of the galaxy formation. 
Aleksander Wolszczan, who discovered the first extrasolar planets, started his 
career at the Copernicus Center in the nineteen-eighties. Professor Miroslaw 
Giersz studied the structure and stability of the globular clusters   systems 
hosting the oldest stars. Currently, Professor Michal Rozyczka and his group are 
developing numerical simulations to study in detail the process of planet 
formation.

Physical cosmology is a relatively young science. It only started after Hubble's 
discovery of the expansion of the Universe, Einstein's formulation of General 
Relativity and the discovery of the microwave background. Yet within only a few 
decades, people were able to measure the size of the Universe, its age, chemical 
composition and put into chronological order major events in cosmic history. 
This process of discovery still continues. In the last few years, observations 
of distant supernovae (exploding stars which for a couple of days outshine their 
host galaxy) at the edge of the observable Universe, and measurements of the 
temperature fluctuations of the microwave relict radiation have proved that the 
Universe is filled with some strange, nonbarionic forms of matter and energy. 
About one third of the whole material content of the Cosmos is in the form of 
dark matter-nonbarionic particles forming massive halos around galaxies and 
clusters of galaxies. About two thirds is in the form of dark energy even more 
mysterious than dark matter causing the acceleration of cosmic expansion.

Many physicists consider dark energy, acting against gravity to accelerate the 
expansion of space, the biggest unsolved problem in physics. It is being 
speculated that dark energy is related to the properties of the physical vacuum. 
Solutions to this problem will most probably require a new, deeper physical 
theory of the fundamental properties of matter. Quite unexpectedly, studies of 
the largest accessible scales the Universe as a whole have revealed something of 
extreme importance about the nature of fundamental processes. The presence of 
dark energy in Cosmos may be the result of processes which took place almost at 
the very BEGINNING of the Universe.

\end{document}